\pdfoutput=1 
\documentclass{sf2a-conf2024}
\usepackage{graphicx}
\usepackage{hyperref}
\usepackage[]{natbib}  
\usepackage{epstopdf}
\usepackage{comment}
\usepackage{mwe} % new package from Martin scharrer
\usepackage{caption}

\def\BibTeX{{\rm B\kern-.05em{\sc i\kern-.025em b}\kern-.08em
    T\kern-.1667em\lower.7ex\hbox{E}\kern-.125emX}}
\bibpunct{(}{)}{;}{a}{}{,}  %%%%%%%%%%%%%  A&A bibliography style
\begin{document}

\TitreGlobal{SF2A 2024}

%%-----------------------------------------------------------------
%%      the top matter
%%

\title{Kilometer-baseline interferometry: science drivers for the next generation instrument}

\runningtitle{Kilometer-baseline interferometery at VLTI}

\author{G. Bourdarot}\address{Max-Planck-Institut f\"ur extraterrestrische Physik, Gießenbachstraße 1, 85748 Garching bei M\"unchen, Germany}

\author{F. Eisenhauer$^{1,}$}\address{Department of Physics, TUM School of Natural Sciences, Technical University of Munich, 85748 Garching, Germany}

%% Keep this line, even if the page will be settled afterwards.
\setcounter{page}{237}

%%-----------------------------------------------------------------

\maketitle

%%-----------------------------------------------------------------
%%        The abstract
%% 
%%  Warning!  within the abstract:
%%  - do not use macros. 
%%  - do not use commands like: \cite, \citet, \citep ... etc.

\begin{abstract}
Infrared interferometry has seen a revolution over the last few years. The advent of GRAVITY+ is about to enable high-contrast observations, all-sky coverage and faint science up to $K_{\mathrm{mag}}=21$, with the implementation on 8m-class telescope of extreme adaptive optics, wide-field observations, and soon laser guide stars, following a long-term vision of technological and infrastructure development at VLTI. This major progress in sensitivity lift a fundamental limitation of infrared interferometry, namely the brightness temperature achievable with this technique down to milli-arcsecond resolution imaging. This change of paradigm is a crucial element for the expansion of current arrays to a facility up to one to ten kilometer baselines. Micro-arcsecond scales imaging in the infrared on thermal objects, reaching the highest angular resolution possible even compared to VLBI, could offer a unique window in observational astronomy for the next generation instrument.

\end{abstract}

%% Insert the keywords (to appear in the ADS indexing)
%% Keywords must be separated by a comma

%%-----------------------------------------------------------------

\section{Sensitivity}
%%-------------------------
\noindent

\begin{minipage}[t]{0.4\textwidth}
The fundamental limit in resolution of any high-angular resolution instrument is ultimately set by the brightness temperature, and therefore by the sensitivity achieved by a given instrument. 
In the case of VLBI, functioning at radio-wavelength, this limit is largely in the non-thermal domain, which fundamentally confines this technique to the study of quasars and the synchroton emission around a few object classes. 
In the O/IR domain, this limit is typically in the range of 1500K to 30,000K, corresponding to thermal objects. 
Therefore, the brighter the star - for a given surface brightness - the bigger in angular size it is ("bright stars tend to be big"). 
Based on the relation between angular diameters and temperature \citep{Kervella2004, Boyajian2012}, one can thus associate the maxium baseline, as defined by $B = \lambda / 2\theta$, assuming a given brightness temperature. This criterion represents the limit for fringe-tracking for a single long-baseline (Fig \ref{fig1}). 
This highlights the fact that for high-angular resolution observations, there is necessarily a trade-off between sensitivity and angular resolution. 
%For an environment with high-extinction or a lower effective brightess temperature (e.g. M-dwarfs), as one gets less photons per element of angular resolution, the maximum baseline reachable diminishes. 

\end{minipage}
%\hfill
\begin{minipage}[t]{0.55\textwidth}
  \begin{flushright}
    \includegraphics[height=10cm]{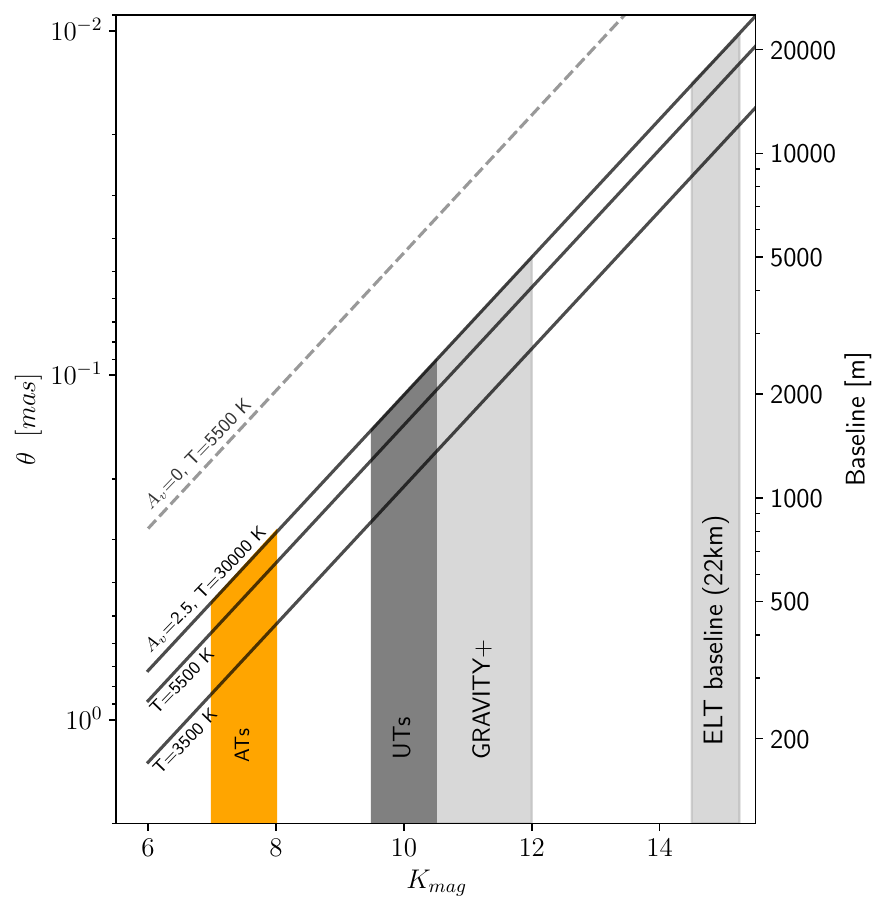}
	\captionof{figure}{Parameter space of infrared interferometery: baseline reachable as a function of sensitivity, computed at $\lambda=2.2\mu m$.}
    \label{fig1} 
  \end{flushright}
\end{minipage}

\noindent In the recent years, a fundamental limit in sensitivity has been overcome with GRAVITY, which now routinely achieves $K_{\mathrm{mag}}=11$ in fringe-tracking (FT) and $K_{\mathrm{mag}}=19$ in science limiting magnitude. For the Galactic Center, the FT limiting magnitude is compatible with baselines up to 5 km assuming a large extinction $A_v$=2.5. For stars located in an environment with no extinction or in the solar neighborhood, the baseline reaches up to 10 km assuming the sensitivity GRAVITY+ on a G-type star.

\begin{comment}
It is worth mentioning that several techniques - such as bootstraping - allow in principle to break this limit. However, they suppose the recombination of multiple baselines in order to reach the same resolution and therefore translates in increased complexity and noise. As shown in Woillez+2017, the excess noise of a bootstraped array increases roughly as $\sqrt{N_{el}}$ for a linear baseline, with $N_{el}$ the number of elements needed to reconstruct a linear baseline. 
This excess noise only decreases in the case of redundancy, such as massively filled imaging array. 
Importantly, to be effective, fringe-tracking with bootstrap must be performed on the closest telescope pairs over which the object is unresolved: splitting the light over more than two channels actually leads in fact to a decrease of SNR (Roddier 1988). 
In the following, we will there keep focusing the discussion on the maximum baseline achievable for a given baseline. These considerations are in any case valid concerning the limit in imaging sensitivity, which is typically of the order of  = 19 ... 21 with GRAVITY+.
\end{comment}

\section{Science drivers}
The $50$-$100\,\mu\mathrm{as}$ imaging resolution and sub-$\mu\mathrm{as}$ astrometry with interferometry would open up a new parameter space in observational astronomy, in a similar fashion to the transformation between single arrays and VLBI in the radio-domain.
In the following, we highlight major science topics that a kilometer array could address.

\begin{figure}[t!]
 \centering
 \includegraphics[width=1.0\textwidth,clip]{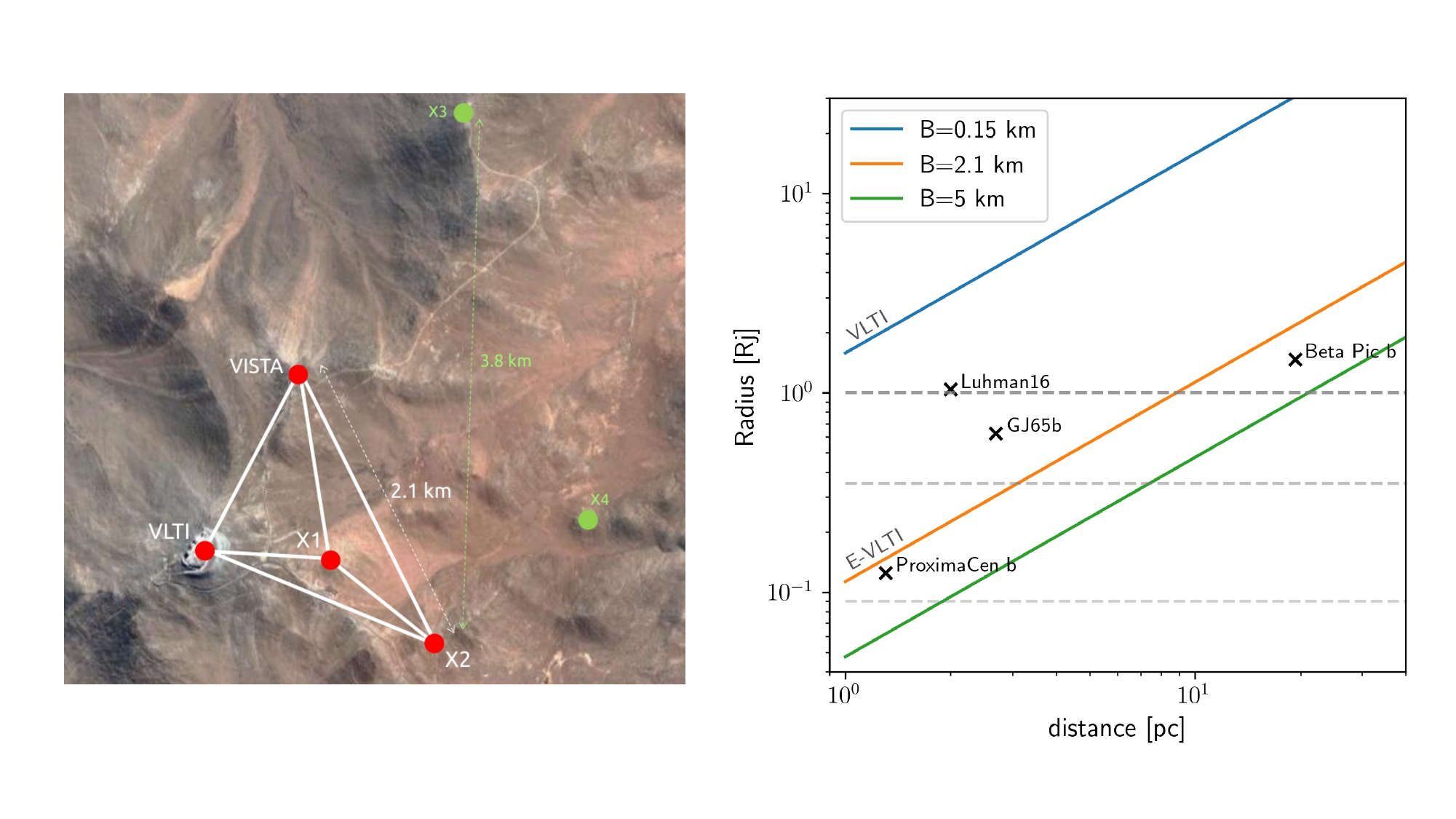}  
%% Note the ABSENCE of the extension .pdf  !
  \caption{a) Potential array layout of an Extended VLTI configuration, using the hills sourrounding the existing infrastructure of Paranal. b) Brown-dwarf and exoplanet radii in our Solar system neighborhood which could be resolved by a kilometer baseline interferometer.}
  \label{fig2}
\end{figure}

\paragraph{Early Universe up to redshift $z=10$} ~\\
The formation of supermassive black-holes (SMBH) and their evolution across cosmic time is a key element in our comprehension of galaxy evolution and black-hole physics. Recent observations with JWST have shown that SMBH are numerous and massive at the earliest stage of the Universe, with a dozen detections of broad line Active Galactic Nuclei (AGNs) between $z=4$ and $10$ \citep{Maiolino2024}. %, and the detection of GN-z11 at $z=10.6$ with a mass $10^6\,\mathrm{M}_{\odot}$.
How do these objects form, are they direct collapse black holes? What is their exact mass, and how do they relate to the formation of SMBH seeds? Recently, GRAVITY+ provided the first dynamical measurement of a black hole mass at redshift $z=2$ \citep{GPLUS2024}, which is the only direct method allowing the measurement of black-hole dynamical masses at high-redshift. Kilometer-wide interferometry, with its $<100\,\mu\mathrm{as}$ imaging resolution and sub-microarcsecond differential astrometry, would probe the mass and the evolution of SMBH up to redshift $z=10$, and allow for full resolution imaging of the BLR region. The imaging capability of a kilometer array could hint at potential changes of its structure at high-redshift and allow the direct imaging of potential binary SMBH. The census of the binary SMBH population and its comparison with LISA \citep{LISA2023} would be a key element to answer the final parsec problem and the coalescence of SMBH. The study of particular objects, such as OJ 287, which is thought to consist of a binary SMBH where the companion is plunging in the accretion disk of the primary with a period of 12 years, offers the opportunity of spatially resolved observations and temporal monitoring over several years. 

\paragraph{Black Hole accretion \& No-Hair theorem}  ~\\
In our own Milky Way, the Galactic Center is a unique laboratory for experimental black-hole studies \citep{Genzel2024}. The monitoring of the stellar orbits of the S-stars with adaptive optics \citep{Gillessen2009,Ghez2005} and infrared interferometry \citep{GRAVITY2017}, allowed the measurement of the black-hole mass with an unmatched accuracy and general relativity tests, including the Schwarzschild precession around SgrA*, relativistic gravitational redshift or the local position invariance tests. 
%In the radio-domain, the EHT interferometry array produced the image of SgrA* environment with $25\,\mu\mathrm{as}$. 
Kilometer baseline interferometer will allow to directly resolve the accretion region of SgrA* and IR flare activity at the innermost circular orbit radius with a resolution $<100\,\mu\mathrm{as}$.
In addition, the astrometric accuracy of interferometry would allow to constrain the black-hole spin and no-hair theorem tests. 
This could already be achieved by combining ELT-MICADO spectroscopy \citep{Davies2021} and an upgraded version of GRAVITY with $<20\mu\mathrm{as}$ astrometry, if an S-star is found with a pericenter 2 to 4 times smaller than S2.
%Lastly, provided with the wide-field capability could be applied to kilometer baseline, one could envision to expand the S2 experiment in a local extragalacic object.

\begin{figure}[b]
 \centering
 \includegraphics[width=1.1\textwidth,clip]{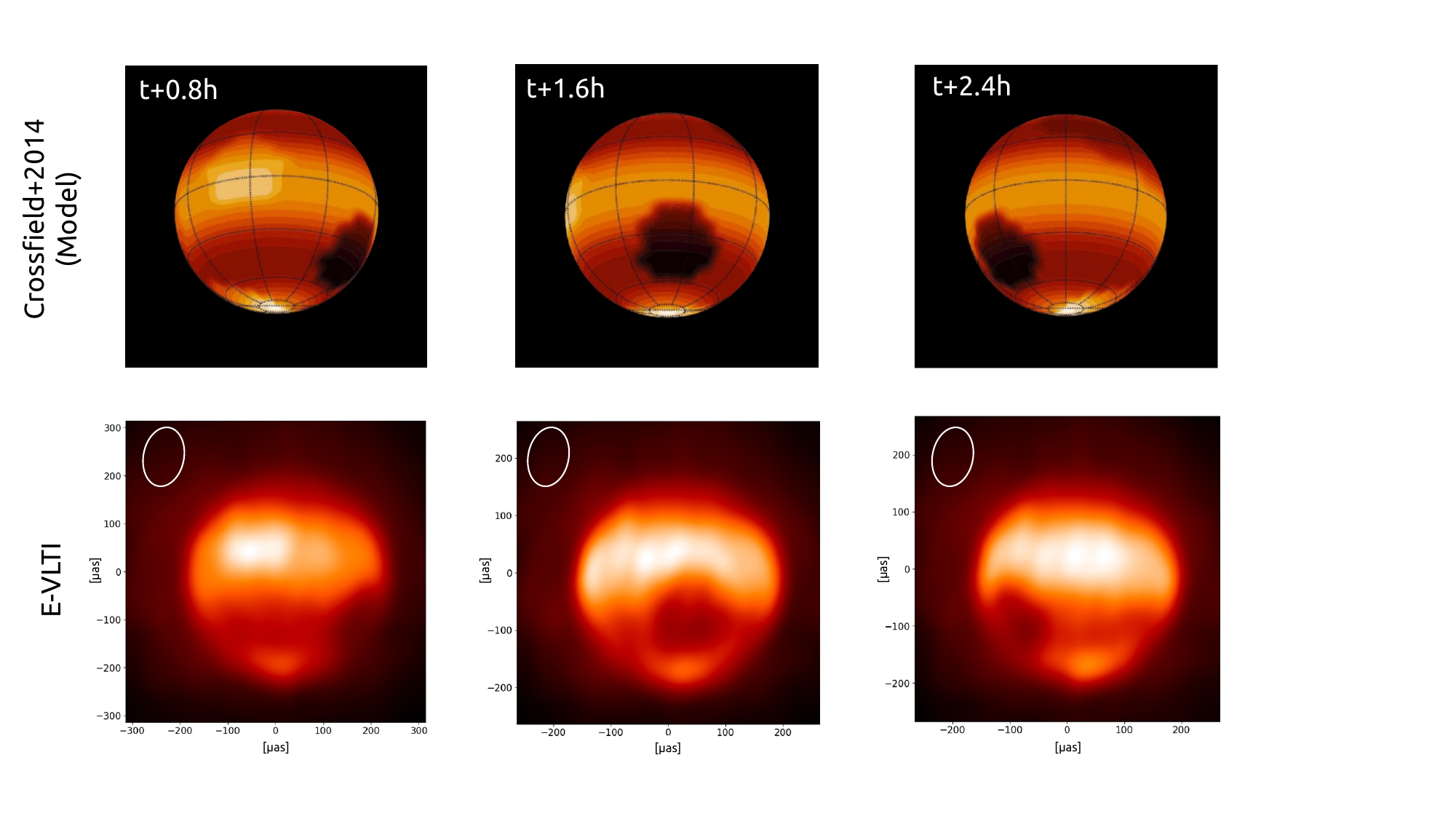}  
%% Note the ABSENCE of the extension .pdf  !
  \caption{Simulated image of Luhman 16B with a E-VLTI, based on the model of \cite{Crossfield2014}.}
  \label{fig3}
\end{figure}

\paragraph{Exoplanet Surface and Clouds} ~\\
In the past two decades, ground-based and space observations have transformed our understanding of exoplanet atmospheres, showing a great diversity in composition and dynamics. Directly imaged brown dwarfs and exoplanets offer a key case for giant planets studies, and are expected to create large atmosphere circulation and clouds due to rapid rotation and strong interior convection. 
Yet, our knowledge of cloud coverage and properties, despite their profund influence on the atmosphere dynamic and radiative transfer, is largely limited. 
Brown-dwarf in particular exhibit patchiness on regional to global scales, as traced with doppler imaging, photometry, or spectroscopic monitoring \citep{Crossfield2014, Biller2024}. %Apai2021,
However, these techniques are intrisically non-resolved and limited by strong spatial degeneracies.
%The direct imaging of the surface itself requires an angular resolution of typically $50$ - $100\,\mu\mathrm{as}$ in the near-infrared, which can only be achieved by kilometer baseline interferometry. 
Kilometer-baseline interferometry is the only technique able to spatially resolve these angular scales, an array with 2km to 5km being able to resolve the surface of brown dwarf and exoplanets in the Solar System neighborhood (Fig \ref{fig2}), including Luhman 16 (Fig \ref{fig3}). 
%A prime target for this study is Luhman 16, for which a 2km-baseline allows about 5 spatial resolution elements across its surface.
These observations also allow the direct measurement of the radius, placing a calibration point on the evolutionary models of brown dwarfs and exoplanets. 
The imaging and astrometry resolution has also proven to constrain the potential presence of satellites in these systems, as recently demonstrated with GRAVITY (Xuan et al., accepted in Nature).
Finally, for young planets, a $100\,\mu\mathrm{as}$ resolution directly constrains the size of the circumplanetary disk (CPDs) in the near-infrared \citep{Rab2019}.
%, in which these satellites are expected to form.

\section{Extended Very Large Telescope Interferometer}
%%---------------------------------
The Very Large Telescope Interferometer (VLTI) located at Paranal Observatory is the most sensitive interferometry facility to this day \citep{Eisenhauer2023}. 
Based on the current sensitivity of GRAVITY+, the addition of telescopes with 4m to 8m diameter is compatible with the recombination of baselines up to 2 km to 5 km (Fig \ref{fig1}), offering the opportunity to leverage the existing infrastructure of the VLTI to larger angular resolution.
The site of Paranal itself and its surrounding infrastructure (VISTA, SPECULOOS) has an array configuration with a few kilometer width (Fig \ref{fig2}).
%In terms of top-level-requirements, the interferometric combiner should provide dual-field capability, to allow for long integration on faint targets, with high-order AO correction.
The recombination would rely on more than three decades of developments in that area, and the development of key technologies, which includes:
\begin{itemize}
\item dual-field observations, to allow for long integration on faint targets. 
\item delay lines of the order of 1m/s speed (up to 10km baseline), and a total stroke covered by double-pass or incremental steps for the DC part.
\item propogation including either optical fiber as explored initially at Mauna Kea in the 'OHANA project \citep{Mariotti1998}, free-space propagation with adaptive optics \citep{Horst2023}, or vacuum pipes \citep{Mozurkewich2016}.
\end{itemize}

\section{Conclusions}
%%--------------------
The major breakthrough of near-infrared interferometry and the current sensitivity in the range of GRAVITY+ era open up a fundamentally new parameter space in high-angular resolution observations. Building upon the existing facility, the current sensitivity would be compatible with the extension of the interferometry facility to kilometer-wide baseline, based on 4m to 8m-class telescopes. Such an array would provide an imaging resolution of the order of 50-100$\mu\mathrm{as}$ in the infrared and sub-$\mu$as astrometric resolution in the near-infrared, addressing a unique domain of fundamental physics and astrophysics. %, from black-holes to exoplanet atmospheres.

% Optional acknowledgements
% -------------------------
\begin{acknowledgements}
G.B. acknowledges the funding of the Max Planck Society through the R.Genzel Nobel Laureate Fellowship. The authors thank S.Gillessen for his careful reading and correction of the manuscript.
\end{acknowledgements}

\bibliographystyle{aa}  % A&A bibliography style file (aa.bst)
\bibliography{Bourdarot_S05} % your references in file: Yourfile.bib

\end{document}